\documentclass[aps,prb,twocolumn,showpacs,superscriptaddress,floatfix,nofootinbib]{revtex4}
\usepackage{amsfonts}

\usepackage{graphicx}
\usepackage{amsmath}
\usepackage{times}
\usepackage{amssymb}
\usepackage{color}
\usepackage[colorlinks,bookmarks=false,citecolor=blue,linkcolor=red,urlcolor=blue]{hyperref}

\begin{document}

\title{
 Competing states in the SU(3) Heisenberg model on the honeycomb lattice:\\
 Plaquette valence-bond crystal versus dimerized color-ordered state  
}

\author{Philippe Corboz}
\affiliation{Theoretische Physik, ETH Z\"urich, CH-8093 Z\"urich, Switzerland}

\author{Mikl\'os Lajk\'o}
\affiliation{Institute for Solid State Physics and Optics, Wigner Research
Centre for Physics, Hungarian Academy of Sciences, H-1525 Budapest, P.O.B. 49, Hungary}
\affiliation{Department of Physics, Budapest University of Technology and Economics, 1111 Budapest, Hungary}

\author{Karlo Penc}
\affiliation{Institute for Solid State Physics and Optics, Wigner Research
Centre for Physics, Hungarian Academy of Sciences, H-1525 Budapest, P.O.B. 49, Hungary}
\affiliation{Department of Physics, Budapest University of Technology and Economics, 1111 Budapest, Hungary}
%
\author{Fr\'ed\'eric Mila}
\affiliation{Institut de th\'eorie des ph\'enom\`enes physiques, \'Ecole Polytechnique F\'ed\'erale de Lausanne (EPFL), CH-1015 Lausanne, Switzerland}
\author{Andreas M. L\"auchli}
\affiliation{Institut f\"ur Theoretische Physik, Universit\"at Innsbruck, A-6020 Innsbruck, Austria}

\date{\today}

\begin{abstract}
Conflicting predictions have been made for the ground state of the SU(3) Heisenberg model on the honeycomb lattice: Tensor network simulations found a plaquette order [Zhao et al, Phys.~Rev.~B {\bf 85}, 134416 (2012)], where singlets are formed on hexagons, while linear flavor-wave theory (LFWT) suggested a dimerized, color ordered state [Lee and Yang, Phys.~Rev.~B {\bf 85}, 100402 (2012)]. In this work we show that the former state is the true ground state by a systematic study with infinite projected-entangled pair states (iPEPS), for which the accuracy can be systematically controlled by the so-called bond dimension $D$. %
Both competing states can be reproduced with iPEPS by using different unit cell sizes. For small $D$ the dimer state has a lower variational energy than the plaquette state, however, for large $D$ it is the latter which becomes energetically favorable. 
%
The plaquette formation is also confirmed by exact diagonalizations and variational Monte Carlo studies, according to which both the dimerized and plaquette states are non-chiral flux states.
\end{abstract}


\maketitle

\section{Introduction}
Systems of fermions with multiple flavors have attracted increasing interest recently thanks to the proposals to experimentally realize such systems with ultra-cold fermionic atoms in optical lattices~\cite{wu2003,honerkamp2004,cazalilla2009,gorshkov2010} and  the rapid experimental progress in the field.~\cite{takasu03, fukuhara07,fukuhara07b,fukuhara09, kraft09, stellmer09,deescobar09,mickelson10,desalvo10,tey10,stellmer10,taie10,stellmer11,sugawa11,stellmer13} In general these systems can be described by a Hubbard model with $N$ flavors (or colors) of fermions.  In the limit of strong on-site repulsion and an integer filling per lattice site, the system is in a Mott insulating state, and the low-energy physics is captured by the SU(N) Heisenberg model. These models give rise to a rich variety of exotic quantum states, such as different N\'eel-type states,~\cite{toth2010, Bauer12, corboz11-su4} generalized valence-bond solids,~\cite{Corboz07,arovas2008,hermele2011,Corboz12_simplex, Song13} algebraic N-flavor  liquids,~\cite{affleck1988-sun,Assaad05,xu2010,Corboz12_su4,Cai12}  chiral spin liquids,~\cite{hermele2009,hermele2011,szirmaiG2011, Song13}, and more.\cite{Rapp08,Manmana11, Rapp11}

The SU(N) Heisenberg Hamiltonian is given by 
\begin{equation} 
\label{eqn:H}
H =  \sum_{\langle i,j \rangle} \sum_{\alpha, \beta} |\alpha_i \beta_j \rangle \langle \beta_i \alpha_j | =  \sum_{\langle i,j \rangle} P_{ij},
\end{equation}
where the first sum goes over nearest neighbors pairs and $\alpha$, $\beta$ run over the $N$ possible colors (flavors) at each site. $P_{ij}$ is a permutation operator which exchanges colors on neighboring sites.  In the present work we will focus on the case of one particle per site (corresponding to the fundamental representation) and $N=3$ on the honeycomb lattice. 

In general the theoretical study of these models is very challenging, particularly because in many cases these models exhibit a negative sign problem in Quantum Monte Carlo simulations, in contrast to the $N=2$ case on bipartite lattices. Therefore, other methods have to be used to study the physics of these models. Mean-field theory typically fails to correctly predict the ground state. In most cases the classical solution exhibits an infinite degeneracy, which is lifted upon including quantum fluctuations. 

One way to go beyond the simple mean-field (Hartree) solution is through linear flavor-wave theory, which takes into account quantum fluctuations on top of a Hartree solution at the harmonic level. This method has successfully accounted for the three-sublattice order in the SU(3) Heisenberg model on the square lattice.\cite{Bauer12} Further quantum fluctuations can in principle be included by taking higher-order terms into account, but this has not been achieved yet.

A powerful class of methods which enable a systematic study of the solution upon adding quantum fluctuations are tensor network algorithms. The most famous method in this class is the density matrix renormalization group (DMRG) method,~\cite{white1992} which is the state-of-the-art method to simulate (quasi-) one dimensional systems. DMRG is based on a variational ansatz called matrix product state (MPS), where the coefficients of the wave function are efficiently encoded by a product of matrices. Substantial progress has also been made in the simulation of two-dimensional systems with extensions of the MPS to higher dimensions, a so-called projected entangled-pair state (PEPS)  or tensor product state.\cite{verstraete2004,Verstraete08} As in an MPS, the accuracy of a PEPS can be controlled by the so-called bond dimension $D$.
   
\begin{figure}
\begin{center}
\includegraphics[width=8.5cm]{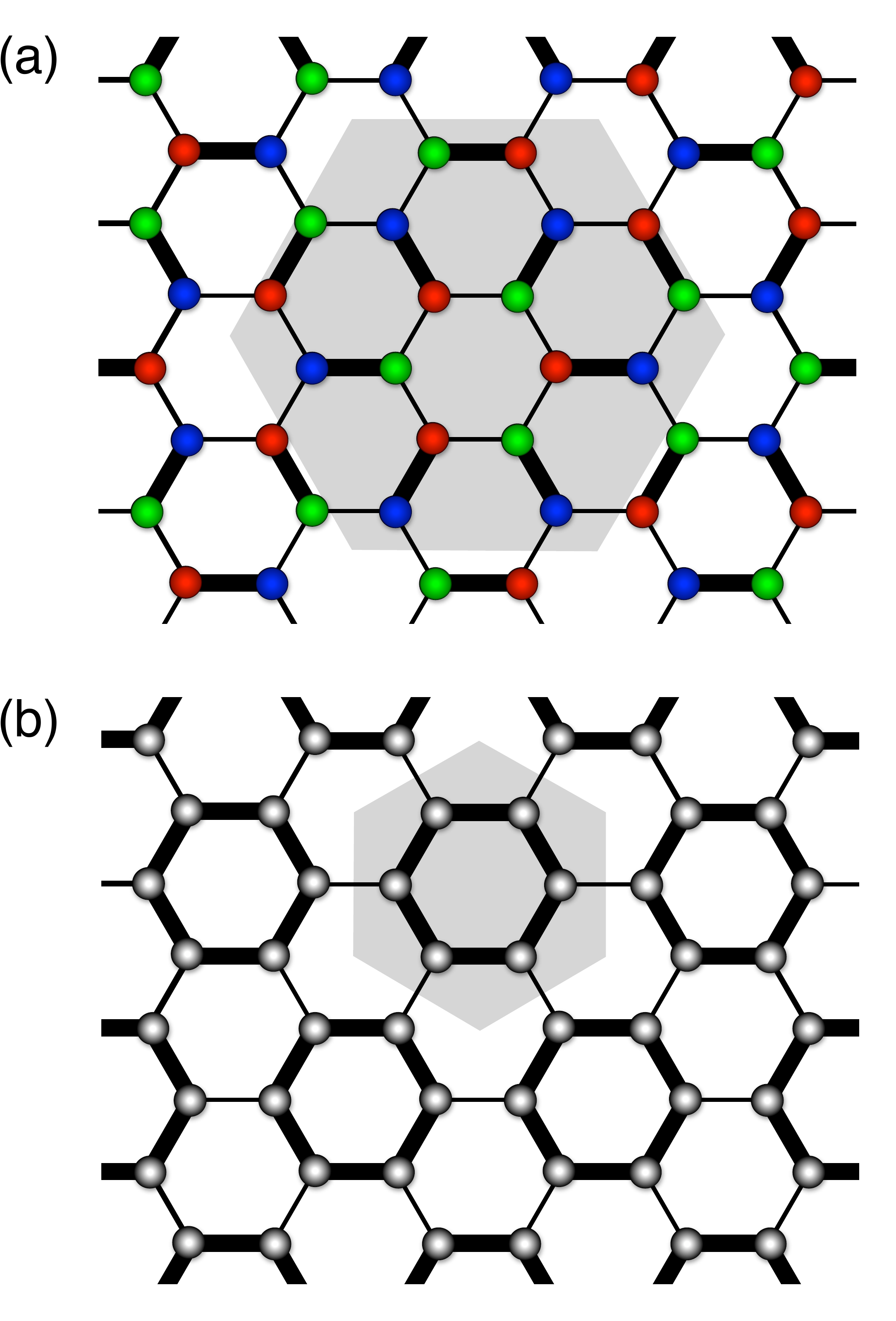} 
\caption{(Color online) Schematic illustration of the two competing  states of the SU(3) Heisenberg model on the honeycomb lattice. Thick (thin) bonds correspond to low (high) bond energies. (a) Color-ordered, dimerized state obtained with a 18-site unit cell (shaded in grey).  
(b) Plaquette state with a 6-site unit cell which has no color order.}
\label{fig:states}
\end{center}
\end{figure}

Based on these two approaches, conflicting results have been reported for the SU(3) Heisenberg model on the honeycomb lattice recently.~\cite{Lee12,Zhao12} In both studies the bilinear-biquadratic spin-1 model has been considered, which includes the SU(3) Heisenberg model as a special point in the phase diagram (when the coefficients of the bilinear and biquadratic terms are equal and positive). LFWT~\cite{Lee12} predicts a dimerized, color-ordered state in a 18-site unit cell, depicted in Fig.~\ref{fig:states}, whereas in the variational tensor network study in Ref.~\onlinecite{Zhao12} a plaquette order has been found by using a 6-site unit cell. LFWT is not a variational method and therefore the energies from the two approaches cannot be directly compared. Furthermore, a 18-site unit cell, which would be compatible with the dimerized state, has not been tested in the tensor network study, and thus it is still an open question what the true ground state is. 

The aim of our work is to unambiguously identify the true ground state of the SU(3) Heisenberg model on the honeycomb lattice by a systematic study of the energetics of the two competing states by means of infinite PEPS (iPEPS) with different unit cell sizes and bond dimensions. We show that the dimerized state predicted by LFWT can be understood as a low-entanglement solution which is reproduced with iPEPS for small bond dimensions. This state, however, is metastable for large bond dimensions, and the true ground state is the plaquette state found in the previous tensor network study. We provide further support for this state also from exact diagonalization results up to system size $N_s=24$. Finally, using Gutzwiller-projected free-fermion wavefunction, we characterize the competing states based on the properties of the fermionic wave function.

The paper is organized as follows: In Sec.~\ref{sec:iPEPS} we provide a brief introduction to iPEPS, in Sec.~\ref{sec:ipepsresults} we present the iPEPS results, in Sec.~\ref{sec:edresults}  the ED results and
in Sec.~\ref{sec:varresults} the variational Monte Carlo (VMC) results from Gutzwiller projected fermionic wavefunctions. Finally we summarize our conclusions in Sec.~\ref{sec:concl}.

\section{Infinite projected entangled-pair state (iPEPS)}
\subsection{Method}
\label{sec:iPEPS}

In this section we give a short summary of the tensor network method used in this work, and point out the differences to the approach used in Ref.~\onlinecite{Zhao12}. For further details on the method we refer to previous works,\cite{Verstraete08,jordan2008,corboz2010} in particular also Ref.~\onlinecite{Corboz12_su4} where we used the same approach for the SU(4) Heisenberg model on the honeycomb lattice.

Each tensor network algorithm has three essential ingredients: (1) the structure of the tensor network ansatz, (2) the optimization method (i.e. how to find the optimal values for the tensors to have an approximate representation of the ground state), and (3) the method used to compute expectation values of observables (i.e. the contraction of the tensor network). 

(1) The tensor network ansatz we use is a projected entangled-pair state (PEPS).\cite{verstraete2004,Verstraete08} It is a variational ansatz aimed at efficiently representing ground states of two-dimensional lattice models. The coefficients of the wave function are obtained by taking the trace of a product of tensors, with one tensor per lattice site. Each tensor has a physical index which carries the local Hilbert space of a lattice site of dimension $d$, and auxiliary indices which connect to the neighboring tensors on the lattice. These auxiliary indices have a certain dimension $D$ which is called the bond dimension. On the square lattice each tensor $T^p_{ijkl}$ has $dD^4$ elements,  whereas on the honeycomb lattice each tensor $T^p_{ijk}$ has $dD^3$ elements. The numbers stored in these tensors are the variational parameters of the ansatz, i.e. the larger $D$ the more variational parameters, and therefore the (potentially) more accurate the ansatz. The special case of $D=1$ corresponds to a product state with a vector $T^p$ on each site. An infinite PEPS (iPEPS) \cite{jordan2008} consists of a unit cell of different tensors, which is periodically repeated on the lattice to represent a state directly in the thermodynamic limit. Using different unit cell sizes enables to represent states with different types of translational symmetry breaking. 

(2) An approximate representation of the ground state is found by performing an imaginary time evolution of a random initial iPEPS. The imaginary time evolution operator $\exp(-\beta \hat H)$ is split into a product of two-body operators via a second order Trotter-Suzuki decomposition (see Ref.~\onlinecite{corboz2010}). Multiplying such a two-body operator to the iPEPS increases the bond dimension of the corresponding bond between the sites the operator is acting on. To constrain the computational cost the auxiliary space of the bond has to be truncated down to the original bond dimension $D$ after a two-body operator has been applied. There are different ways to perform this truncation, as discussed e.g. in Ref~\onlinecite{corboz2010}. With the \textit{full update} a bond is truncated in an optimal way by taking into account the whole wave function to find the relevant subspace. The \textit{simple update} \cite{vidal2003-1,jiang2008} is computationally cheaper since it involves only local tensors surrounding the bond to be truncated, however, it is not optimal. In the present work we use the more accurate full update (in contrast to the previous study in Ref.~\onlinecite{Zhao12}).

(3) To evaluate observables the tensor network needs to be contracted (by computing the trace of the product of all tensors). This contraction can only be done approximately in polynomial time. As in previous works we use the corner transfer matrix method~\cite{nishino1996, orus2009-1} generalized to arbitrary unit cell sizes.~\cite{corboz2011} We map the honeycomb lattice onto a brick-wall square lattice as explained in Ref.~\onlinecite{Corboz12_su4}. We checked that quantities of interest are converged in the "boundary" dimension $\chi$, which controls the accuracy of the contraction. 

In order to reduce the computational cost we use tensors with $\mathbb{Z}_q$ symmetry, which is a discrete subgroup of SU(3).\cite{singh2010,bauer2011} The tensors then acquire a block structure, similar to a block diagonal matrix.

\begin{figure}
\begin{center}
\includegraphics[width=8.5cm]{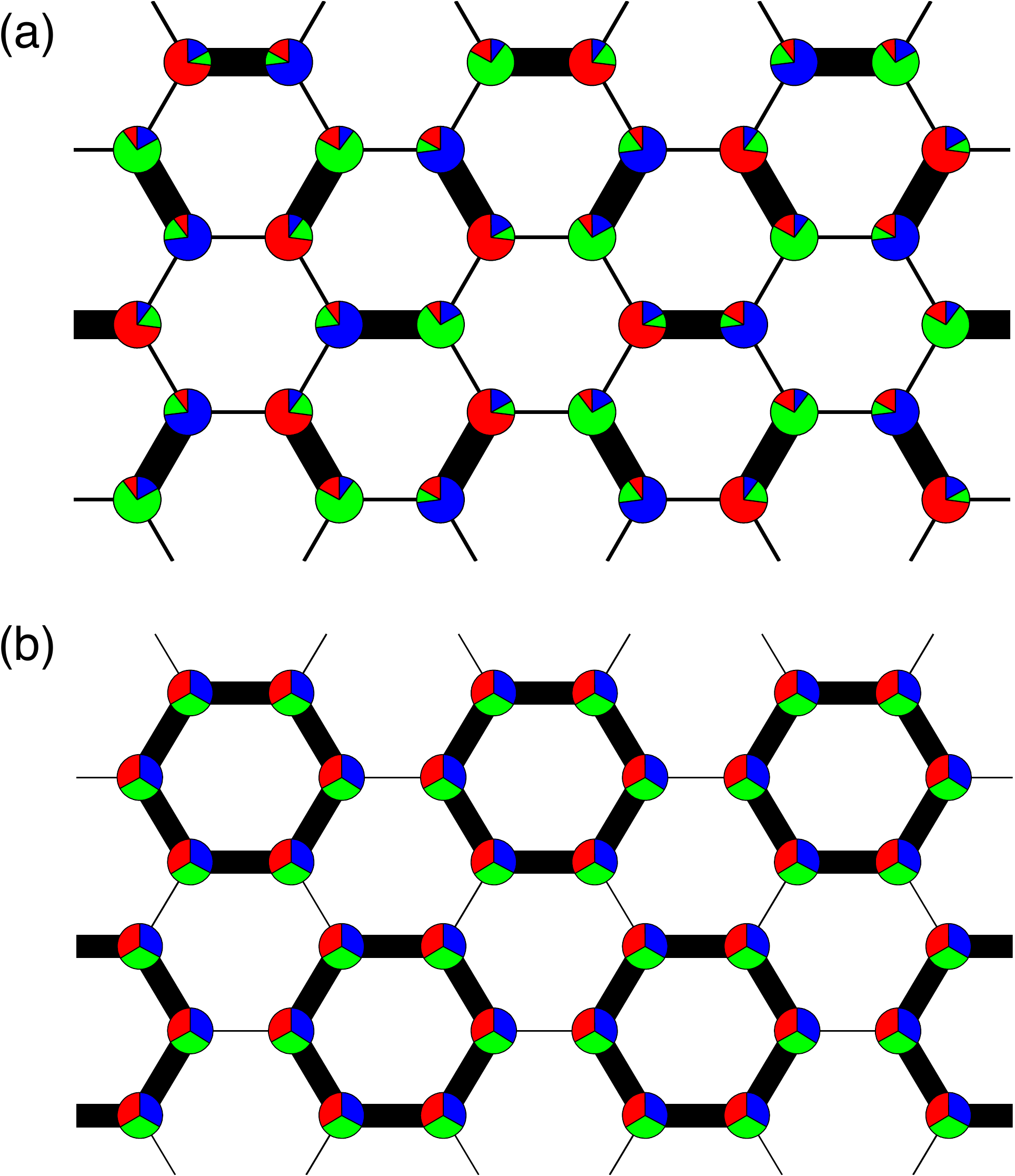}
\caption{(Color online) Local order parameters obtained with iPEPS obtained with two different unit cell sizes. The thickness of the bonds is proportional to the square of the energy of the corresponding bond. Each pie chart on a site visualizes the local color density. 
(a) The color-ordered, dimerized state obtained with a 18-site unit cell for $D=2$.
(b) Plaquette state which has no color order: Each color has the same density on each site ($D=16$).
}
\label{fig:statesiPEPS}
\end{center}
\end{figure}

\begin{figure}
\begin{center}
\includegraphics[width=8.5cm]{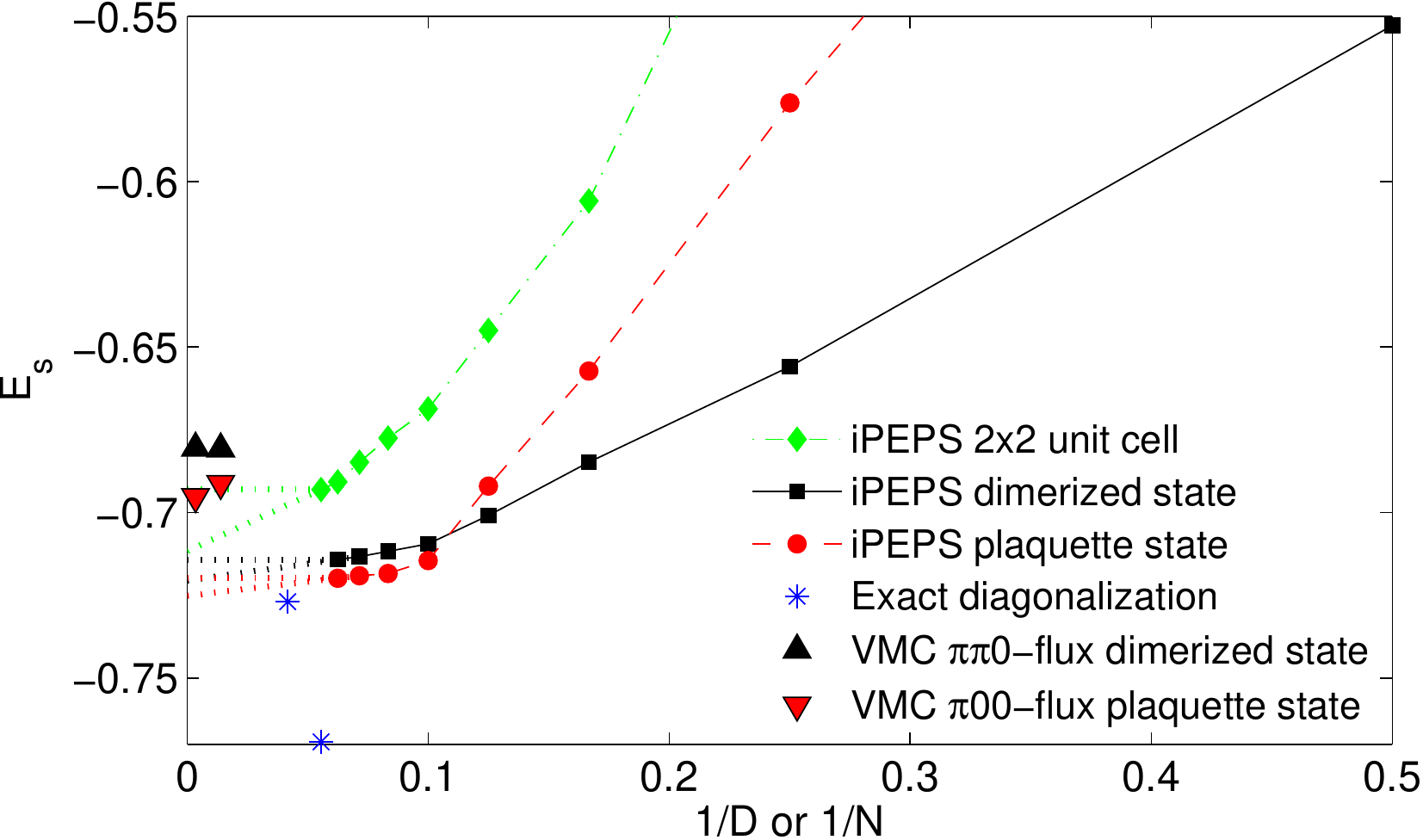} 
\caption{(Color online) iPEPS simulation results for the variational energy as a function of inverse bond dimension $1/D$ for different unit cell sizes, compared to the exact diagonalization (ED) results and the lowest energy states from variational Monte Carlo (VMC). For bond dimensions $D\le8$ the color-ordered, dimerized state has the lowest variational energy, for larger $D$ the plaquette state becomes energetically favorable. The dotted lines are a guide to the eye. 
 }
\label{fig:energy}
\end{center}
\end{figure}

\subsection{IPEPS results}
\label{sec:ipepsresults}

Here we present a systematic study of the solution for the ground state as a function of the bond dimension $D$ in iPEPS, which controls the accuracy of the ansatz, and also the amount of quantum fluctuations (or entanglement) taken into account, as we explain in the following. We consider results for different unit cell sizes: $2\times2$, and the 6-site and the 18-site unit cell shaded in grey in Fig.~\ref{fig:states}(a-b).

A $D=1$ iPEPS corresponds to a product state (a site-factorized wave function), i.e. a non-entangled state. The energy per site is $E_s=0$, which can be easily verified analytically. The state is infinitely degenerate: for example all possible coverings where two nearest-neighbor sites exhibit different colors have the same energy (or more generally, the energy is minimized if the states on neighboring sites are orthogonal). 

With $D=2$ short-range quantum fluctuations are taken into account. It turns out that they lift the infinite degeneracy. If a 18-site unit cell is used in iPEPS, the quantum fluctuations select the same state as predicted by LFWT~\cite{Lee12}: the dimerized, color-ordered state shown in Fig.~\ref{fig:states}(a). In Fig.~\ref{fig:statesiPEPS}(a) we visualize different local quantities obtained from the iPEPS: The thickness of the bonds is proportional to the square of the corresponding bond energy, and the pie charts show the local color density of each color. On each site one of the colors is dominant, and the pattern of the dominant colors  matches the one shown in Fig.~\ref{fig:states}(a). The state is clearly dimerized: the two sites in each dimer have different colors (e.g. green and red), and each dimer is surrounded by four sites where the third color is dominant (e.g. blue). The energy per site is $E_{D=2}=-0.553$. 

If we take smaller unit cells this state cannot be represented, and the  variational energy is  higher, e.g. $-0.421$ in the $2\times 2$ unit cell and $-0.362$ in the 6-site unit cell (using $\mathbb{Z}_q$ symmetric tensors). 

By further increasing $D$ the variational energies in the different unit cells are lowered, as shown in Fig.~\ref{fig:energy}. The energy obtained with the 6-site unit cell decreases more rapidly with  $D$ than the energy obtained with the 18-site unit cell, and it becomes lower for $D\ge10$. The state obtained in the 6-site unit cell is the plaquette state, shown in Figs.~\ref{fig:states}(b) and \ref{fig:statesiPEPS}(b), where low energy bonds are formed around hexagons. On all sites each color has the same density, i.e. the state does not exhibit color-order. This state has already been found in Ref.~\onlinecite{Zhao12}. 
So, the dimerized state is not the true ground state, but only a metastable state which appears when some, but not all of the quantum fluctuations are taken into account. 
We can call it a low-entanglement solution, since it is energetically favorable for small values of $D$ (e.g. $D=2$), at which the iPEPS represents only a slightly entangled state. 


By extrapolating the energy of the plaquette state to the infinite $D$ limit, we expect the ground state energy to lie in between $-0.723$ and $-0.720$. From the slopes of the curves we do not expect another crossing at larger $D$.

\begin{figure}
\begin{center}
\includegraphics[width=0.9\linewidth]{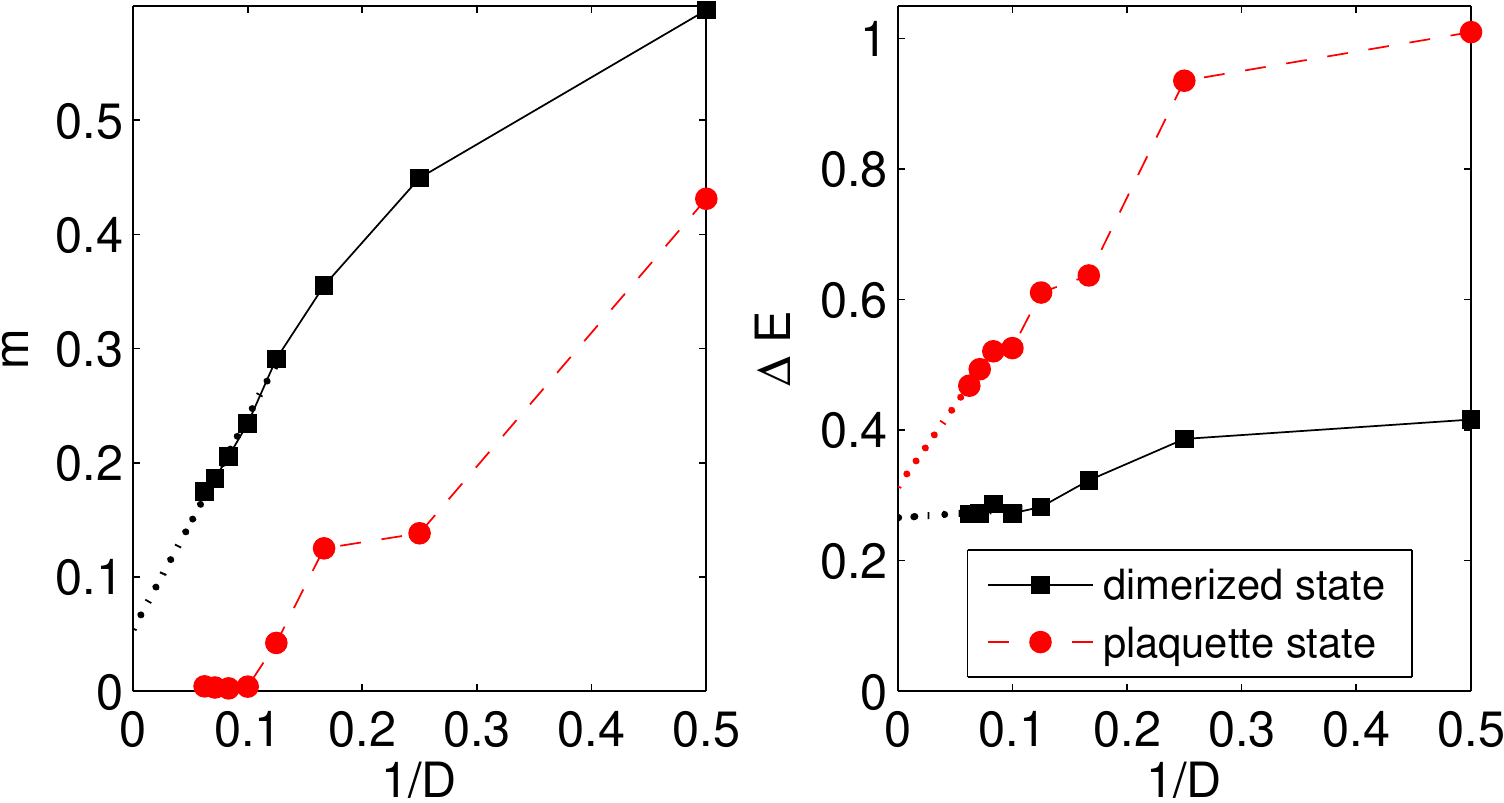} 
\caption{(Color online) iPEPS simulation results of order parameters as a function of inverse bond dimension $1/D$ different unit cell sizes. Dotted lines are a guide to the eye. (a) Local moment which is  suppressed with increasing bond dimension. For the plaquette state it vanishes for $D\ge10$, indicating absence of color-order. Lower right panel: Difference in energy between the strongest bond and the weakest bond in the unit cell, showing that the plaquette state is stable in the infinite $D$ limit. }
\label{fig:OPs}
\end{center}
\end{figure}

In Fig.~\ref{fig:OPs}(a) we show results for the local ordered moment $m$, given by
\begin{equation}
\label{eq:m}
 m=\sqrt{\frac{3}{2} \sum_{\alpha,\beta} \left(\langle S_\alpha^\beta \rangle - \frac{ \delta_{\alpha\beta}}{3 } \right)^2},
\end{equation}
where $S_\alpha^\beta = |\alpha\rangle\langle \beta|$ are the generators of SU(3) and $\alpha,\beta$ run over all flavor indices. 
For the plaquette state $m$ vanishes for $D\ge10$ (i.e. there is no color order). For the dimerized state $m$ remains finite for all values of $D$, however, $m$ is strongly suppressed with increasing $D$. Extrapolating $m$ in $1/D$ yields a finite value, however, since the extrapolated value is small it is difficult to conclude whether the SU(3) symmetry is broken in the infinite $D$ limit or not.

In Fig.~\ref{fig:OPs}(b) we plot the difference between the highest bond energy and the lowest bond energy in the unit cell, 
\begin{equation}
\Delta E = \max(E_b) - \min(E_b),
\end{equation}
which measures the strength of the dimerization or plaquette order. The plaquette order is suppressed with increasing $D$, however, it seems to tend to a finite value in the infinite $D$ limit, which shows that the ground state indeed has long-range plaquette order in the thermodynamic limit. 

We conclude this section with a remark: The plaquette state is compatible with the 18-site unit cell, thus the simulations with this unit cell for $D>8$ should in principle yield the plaquette state. The reason why we remain in the dimer state for $D>8$ is due to metastability. Since the states for lower $D$ are used as an initial state for simulations at larger $D$ one has to overcome an energy barrier to get from the dimer state into the plaquette state when moving from $D=8$ to $D=10$. This does not seem to occur, at least not in the simulated time scales, i.e. the simulation is stuck in the metastable dimer state. We can exploit this fact to compare the energies of the two states at large $D$. 

\section{ED results}
\label{sec:edresults}
In order to corroborate the iPEPS findings we also performed exact diagonalizations
of the SU(3) Heisenberg model on two clusters consisting of $N_s=18$ and $N_s=24$ sites.
Both clusters are compatible with the plaquette state, however the $N_s=18$ is particular in
that it contains additional loops of length six which wrap around the torus. As a consequence 
this cluster allows for more "plaquette" states than clusters with larger circumference. 
Regarding the magnetic dimer state, only the $N_s=18$ cluster is compatible with this type 
of order.

We first calculated the ground state energy per site, which is shown together with the energies
of the iPEPS and the VMC approaches in Fig.~\ref{fig:energy}. The energy per site of the
$N_s=18$ cluster is $-0.76928$, while the value for $N_s=24$ is $-0.72687$, very close to
the large $D$ value inferred from the iPEPS results.
\begin{figure}
\begin{center}
\includegraphics[width=0.8\linewidth]{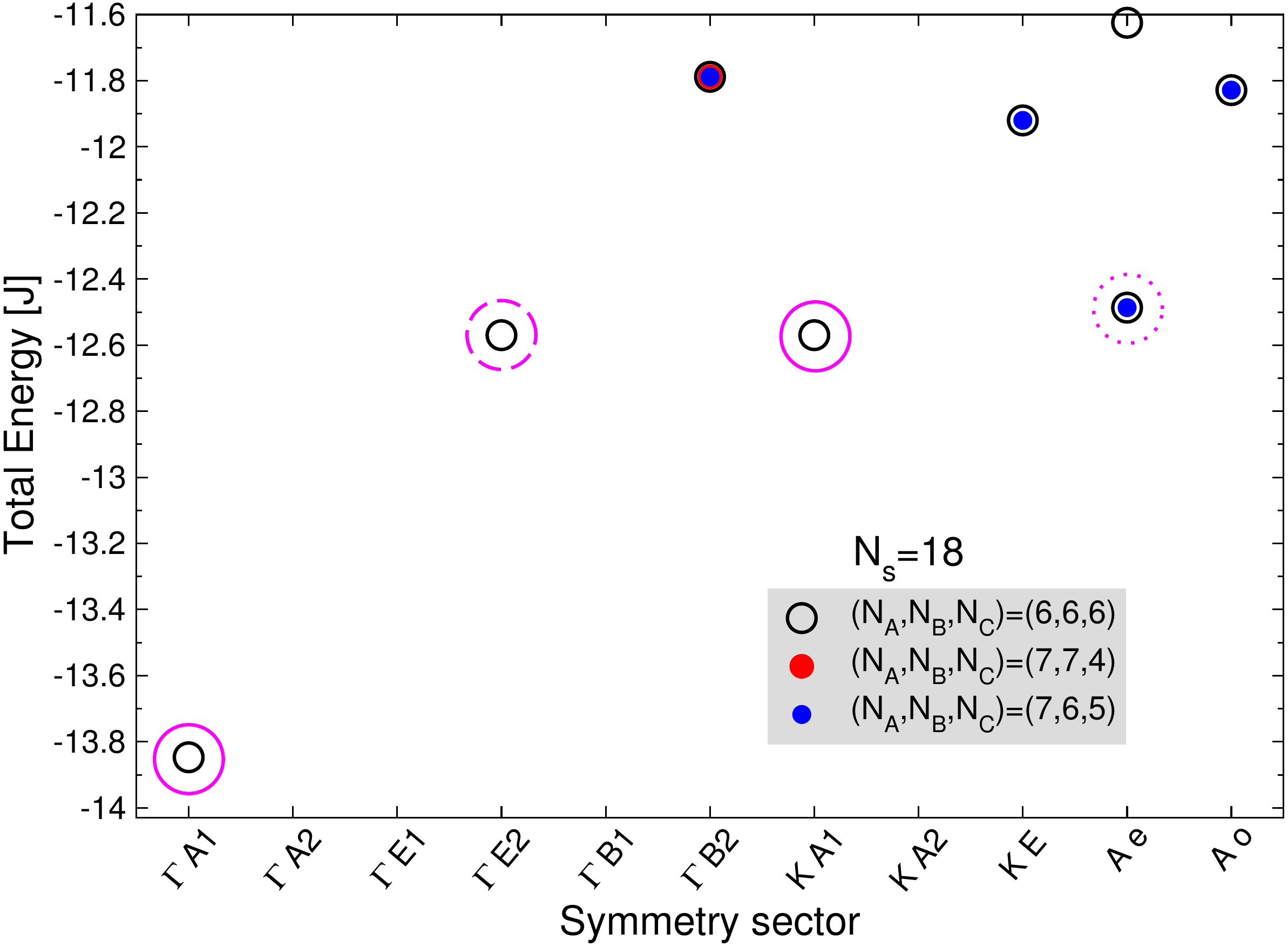} 
\vspace{5mm}$\mbox{}$
\includegraphics[width=0.8\linewidth]{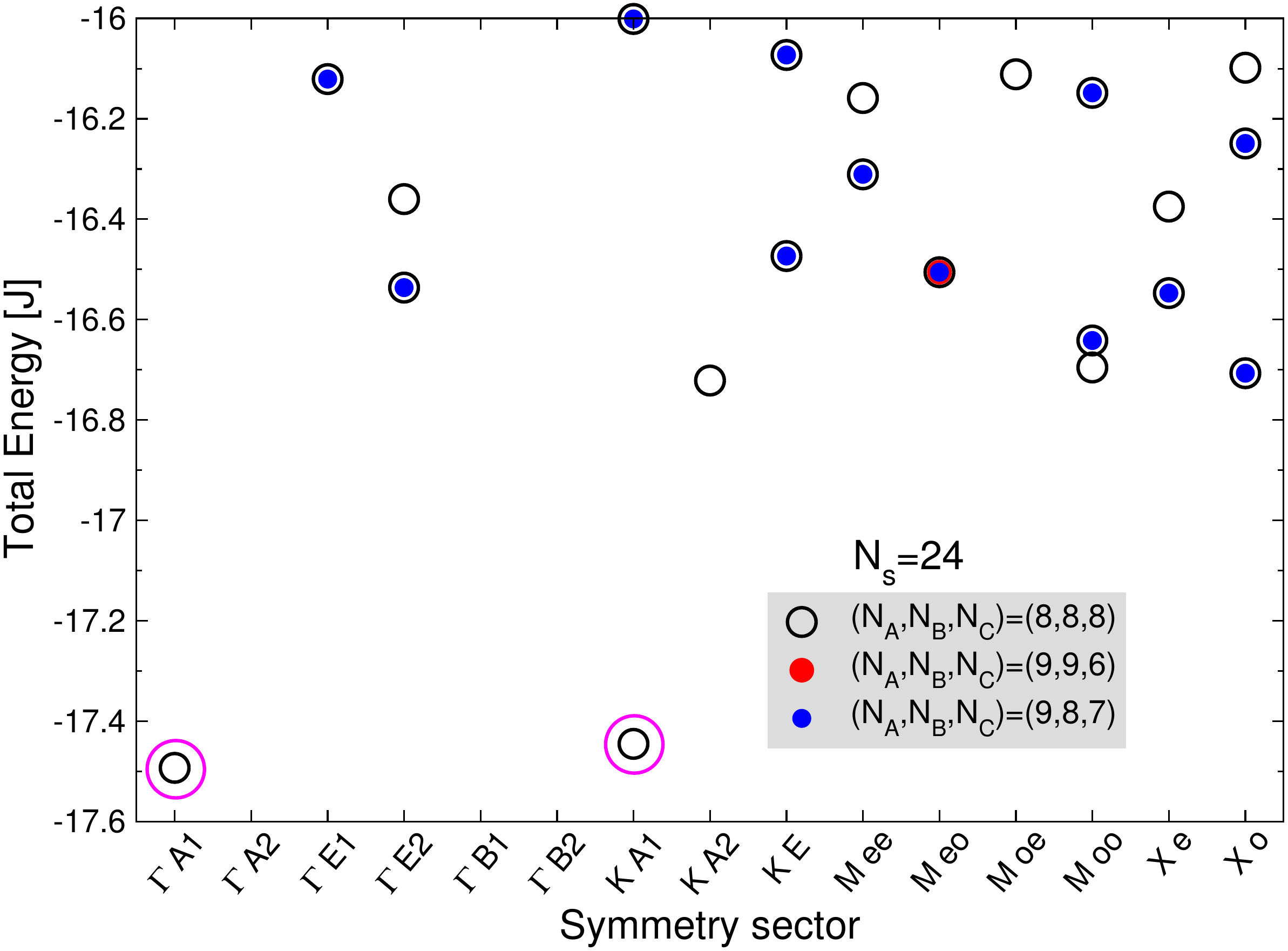} 
\caption{(Color online) Low energy spectra from ED for $N_s=18$ (top figure) and
$N_s=24$ (bottom figure). See main text for explanation of the different symbols.}
\label{fig:ED_Spec}
\end{center}
\end{figure}

Then, we determined the low energy spectrum resolved according to spatial quantum numbers
and for different magnetization sectors, i.e. different values of the three colors $(N_A,N_B,N_C)$. These
spectra are displayed in Fig.~\ref{fig:ED_Spec}. The spectrum of the $N_s=18$ cluster (top panel) 
shows a significant gap between the non-degenerate ground state and a set of states at an energy of about $\sim -12.6$ (encircled 
levels). The two levels highlighted with full line circles are precisely the two levels which are expected for a 
non-magnetic plaquette state with a three-fold ground state degeneracy. 
The (two-fold degenerate) level highlighted with a dashed circle is an artifact of the enhanced 
spatial symmetry of the $N_s=18$ honeycomb cluster. Finally the magnetic level encircled with a dotted circle
would belong to a hypothetical tower of state structure for the magnetic ordering of the dimer state. However
since there is no clear ordering between the levels making up the plaquette state or the dimer state, 
the $N_s=18$ cluster does not seem to be helpful in deciding between the two competing states. 

The situation is much more clear on the $N_s=24$ cluster, where there is an almost perfect degeneracy between the
non-degenerate ground state ($\Gamma$ A1 representation) and the twofold degenerate level in the sector $K$ A1,
fully consistent with a plaquette-like spatial symmetry breaking.

We have also calculated the color-color correlation functions in the ground states of both clusters (not shown), 
and even though the correlations on the $N_s=18$ cluster are consistent in sign with the dimerized magnetic state, 
the actual correlations are quite weak beyond the nearest neighbor correlations, and thus do not lend strong support 
for a magnetically ordered state.

\begin{figure}
\begin{center}
\includegraphics[width=0.9\linewidth]{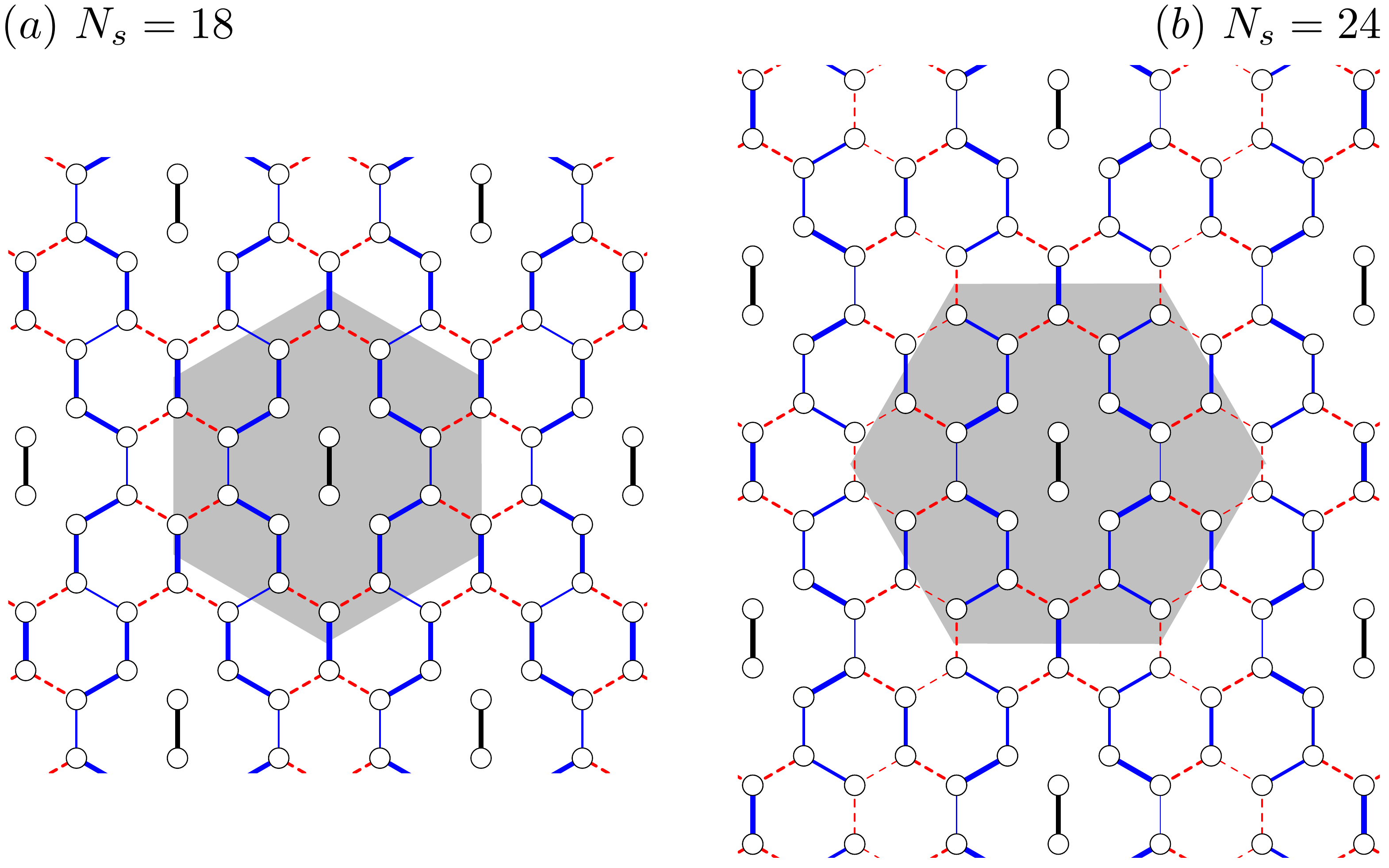} 
\caption{(Color online) Real space bond energy correlation $\langle P_{ij}P_{kl}\rangle-\langle P_{ij}\rangle\langle P_{kl}\rangle$ obtained from ED on clusters of
(a) $N_s=18$ and (b) $N_s=24$ sites. The value of the correlation is proportional to the plotted 
bond strength. Positive (negative) correlations are shown as full blue (red, dashed) lines. The reference
bond is denoted as the bold black bond.}
\label{fig:ED_P2Corrs}
\end{center}
\end{figure}

Finally we show the real-space bond-energy correlations in Fig.~\ref{fig:ED_P2Corrs}. Common to both clusters
is the fact that these correlations extend through the entire cluster, i.e. the correlations are quite long ranged. When it
comes to the signs of the correlations, it turns out that the $N_s=18$ correlations suffer from an admixture of the additional
plaquette states, which are an artifact of this cluster. The correlations in the ground state of the $N_s=24$ cluster do not
suffer from this and most bonds match the expectations of a plaquette ordered singlet state (see e.g. Ref.~\onlinecite{Albuquerque11}
for a discussion of the real-space correlations of $SU(2)$ plaquette  states on the honeycomb lattice). 

In conclusion the ED results are consistent with the plaquette ordering scenario put forward by the Tensor Network and the
iPEPS approach, even though they cannot strictly rule out the magnetic ordering scenario due to the lack of larger clusters which
are compatible with both competing states, and which would allow an unbiased comparison.

\section{Variational Monte Carlo}
\label{sec:varresults}

\subsection{Introduction}

The fermionic representation of the different colors has proven to be a valuable tool to understand the ground state properties of the SU(2) and SU(N) Heisenberg models. The permutation operator $P_{i,j}$ that exchanges the colors between  sites $i$ and $j$ can be written as
\begin{equation}
  P_{i,j} = -\sum_{\alpha,\beta=1}^N f^\dagger_\alpha(i) f^\dagger_\beta(j)
  f^{\phantom{\dagger}}_\beta(i) f^{\phantom{\dagger}}_\alpha(j)  
\end{equation}
in the fermionic representation, where $f^\dagger_\alpha(i)$ creates a fermion of color $\alpha$ at site $i$, and $f^{\phantom{\dagger}}_\alpha(i)$ annihilates it. To treat the four fermion term, it is customary to introduce a bond mean-field decoupling of the $P_{i,j}$ of the form 
\begin{equation}
  P_{i,j}^{\text{MF}} = 
   |\chi_{i,j}|^2 
  -\sum_{\alpha=1}^N
  \left[
    \chi_{i,j} f^\dagger_\alpha(i) f^{\phantom{\dagger}}_\alpha(j)  +
    \chi_{i,j}^* f^\dagger_\alpha(j) f^{\phantom{\dagger}}_\alpha(i)    
    \right] 
    \;,
    \label{eq:PijMF}
\end{equation}
so that $\mathcal{H}^{\text{MF}} = \sum_{\langle i ,j \rangle} P_{i,j}^{\text{MF}}$ describes free fermions.
The hopping amplitude $\chi_{i,j}$ is determined self--consistently as the expectation value
\begin{equation}
    \chi_{i,j} = \sum_{\alpha=1}^N \langle f^\dagger_\alpha(j) f^{\phantom{\dagger}}_\alpha(i) \rangle
\end{equation}
taken in the ground state of the mean-field Hamiltonian $\mathcal{H}^{\text{MF}}$. This approach has been initiated by Affleck and Marston in Ref.~\onlinecite{affleck1988-sun} and has been used both in SU(2) and SU(N) models. Quite interestingly, in some of the mean-field solutions, the product $\prod \chi_{i,j}$ around a plaquette is a complex number $\propto e^{i \Phi}$, as if the fermions were picking up a phase due to a magnetic field with flux $\Phi$ threading through the plaquette (in convenient units). The finite flux can considerably change the band structure of the hopping hamiltonian as well as the correlations. 

 
 In particular, Hermele {\it et al.} in Ref.~\onlinecite{hermele2009} pointed out  that time--invariance breaking chiral solutions with a uniform flux $\Phi$ are good ground state candidates in a particular large-N limit on the square lattice. 

On the honeycomb lattice, the mean-field method has been used for the SU(6) symmetric Heisenberg model in Ref.~\onlinecite{szirmaiG2011}, where several mean-field solutions have been put forward as candidates for the ground state. The lowest mean--field energy solution turned out to be a chiral one, with a finite flux $\Phi=2\pi/3$ per hexagon, in line  with the ideas put forward in Ref.~\onlinecite{hermele2009}. Apart from that, hexamerized solutions with real $\chi_{i,j}$ hoppings were also found.
 While the mean-field solutions give a very useful insight into the possible nature of the ground state, they describe free fermions where the on--site occupancy is fixed to integer (one if the fundamental representation of the SU(N) is considered) on the average only. The treatment of the charge fluctuations beyond mean field is quite involved.
   
There is a complementary approach in which one does not search for mean--field solutions, but one takes the ground--state wave function of some free-fermion Hamiltonian, and, by applying a Gutzwiller projection, one ensures that the occupation of each site is exactly one. The projection is done numerically, using Monte-Carlo importance sampling, where one can efficiently sample the wave function and calculate the energy and correlations.\cite{yokoyama1987,gros1989} For SU(N) Heisenberg chains this approach gives excellent energies and even the correlation functions are reproduced with high accuracy.\cite{Kaplan1982,paramekanti2007,wang2009} Regarding two--dimensional Heisenberg models, the method has been applied to the SU(4) model on the square\cite{wang2009} and honeycomb lattice,\cite{Corboz12_su4} and to the SU(3) model on the triangular lattice.\cite{Bieri2012}
As for the mean--field solution, introducing a nonzero flux for the elementary plaquettes of the hopping Hamiltonian can drastically change the band structure, and, after Gutzwiller projection, lead to an energy lower than that of a purely 0--flux state. This happens for instance in the case of the SU(4) model on the honeycomb lattice,\cite{Corboz12_su4} where introducing a $\pi$--flux leads to an algebraic color liquid that breaks neither the space--group nor the SU(4) symmetry. The energy of the $\pi$--flux state is considerably lower than that of the $0$--flux state, in which the hoppings are all equal. Introducing the $\pi$--flux leads a band--structure with a Dirac-node at the Fermi-level for the quarter filled bands (for $N$ colors of the SU(N) model the filling is $1/N$).

\subsection{Candidate ground states}

\begin{figure}
\begin{center}
\includegraphics[width=8truecm]{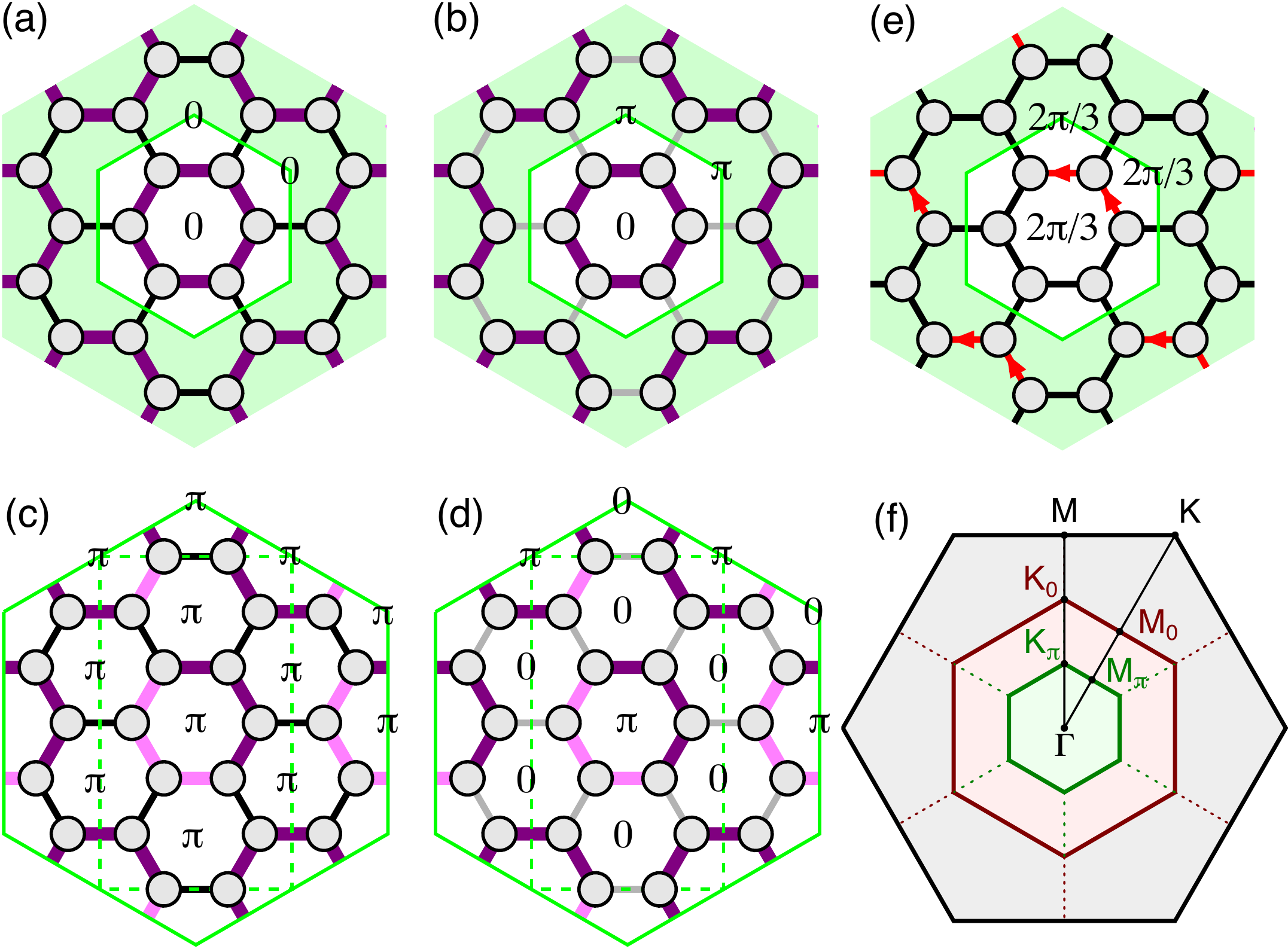} 
\caption{(Color online) 
(a) The $000$--flux and (b) the $0\pi\pi$--flux {\it Kekul\'e} states, with 6 sites and 3 hexagons in the unit cell. (c) The $\pi\pi\pi$--flux and (d) the $\pi00$--flux  states have 12 sites in the primitive unit cell (shown by dashed rectangle) or 24 sites in the hexagonal unit cell.
 These states are characterized by two different absolute values of hopping amplitudes, $t_d$ and $t_h$. The hopping amplitudes on the thin black and grey bonds are $t_d$ and $-t_d$, while on the thick dark and light purple bonds the hopping amplitudes are $t_h$ and $-t_h$, respectively. 
 If the signs of all the hoppings are equal, the product of the hoppings around any of the hexagons is positive: this is the $000$--flux state shown in (a). In the other cases, the product of the hoppings around some of the plaquettes is negative, corresponding to a $\pi$-flux on these plaquettes.
(e) The chiral $\Phi\Phi\Phi$--flux state with $\Phi=2\pi/3$. The red bonds with arrows denote hoppings $t_h e^{ i 2\pi/3}$ with complex amplitudes. Here one can also introduce the {\it Kekul\'e} modulation for the hoppings by changing the sign on the bonds crossing the boundaries of the unit cell ($t_d$ bonds), resulting in a $\Phi \Phi' \Phi'$ flux configuration, with $\Phi' =5 \pi/3$. 
Also $\Phi' \Phi' \Phi'$ and $\Phi' \Phi \Phi$ flux configurations can be created by introducing complex hoppings to the $\pi\pi\pi$ and $\pi 0 0$ case. 
(f) Brillouin zone of the honeycomb lattice (black hexagon) with the high symmetry points $\Gamma=(0,0)$, ${\mathsf K} = (2\pi/ 3\sqrt{3},2\pi/3)$, and ${\mathsf M} = (0,2\pi/3)$. The Brillouin-zone of the $000$, $0\pi\pi$, $\Phi\Phi\Phi$, and $\Phi\Phi'\Phi'$--flux states is shown by the dark red hexagon, with the high symmetry points ${\mathsf K}_0 = (0,4\pi/9)$ and ${\mathsf M}_0 = (2\pi / 3\sqrt{3},2\pi/6)$. The dark green hexagon stands for the Brillouin-zone of the $\pi\pi\pi$, $\pi00$, $\Phi' \Phi' \Phi'$, and $\Phi' \Phi \Phi$--flux states with ${\mathsf K}_\pi = (0,2\pi/9)$ and ${\mathsf M}_\pi = (\pi/3\sqrt{3}, \pi/6)$. The index in ${\mathsf M}$ and ${\mathsf K}$ refers to the flux of the central hexagon realized by real hopping amplitudes. The nearest-neighbor distance is chosen to be unity. 
\label{fig:6and24uc}}
\end{center}
\end{figure}

Since the family of flux states is infinite, it is of course impossible to make a systematic study, and one has to make choices 
guided by simplicity, intuition, or previous results obtained on similar models. In the present case, we have decided to concentrate
on three types of states:
\begin{enumerate}
\item
The two {\it Kekul\'e}-like states\cite{ChangYuHou2007} with $0$ flux in a central hexagon and flux $0$ or $\pi$ in 
the adjacent hexagons called respectively
$000$ and $0\pi\pi$--flux states. These states are compatible with a 6-site unit cell. In this gauge, the hopping amplitudes around the central hexagon are set to $t_h$, while they alternate between $t_h$ and $t_d$ as one goes around the two remaining hexagons in the unit cell, as shown in Figs.~\ref{fig:6and24uc}(a)-(b). The motivation to study these states comes from the on--site and bond color correlations reported in the tensor network simulations\cite{Zhao12} and in the linear flavor-wave theory,\cite{Lee12} as well as our iPEPS calculation (Fig.~\ref{fig:statesiPEPS}) and ED results.
\item
The two  {\it Kekul\'e}-like states with $\pi$ flux in a central hexagon and flux $0$ or $\pi$ in the adjacent hexagons called respectively
$\pi 00$ and $\pi\pi\pi$--flux states.  The realization of these states requires a larger unit cell, as shown in Figs.~\ref{fig:6and24uc}(c)-(d). 
These states are motivated by the results of the SU(4) case on the honeycomb lattice.\cite{Corboz12_su4}
\item 
The uniform chiral $\Phi\Phi\Phi$--flux state with $\Phi=2\pi/3$ per hexagon (Fig.~\ref{fig:6and24uc}(e)), following the mean--field results for the SU(6) Heisenberg model on the honeycomb lattice,\cite{szirmaiG2011,hermele2009} as well as the uniform $\Phi'\Phi'\Phi'$--flux state, where $\Phi'= 5 \pi/3$. Both uniform flux states can be modulated to achieve a $\Phi \Phi' \Phi'$ and a $\Phi'\Phi\Phi$ flux configuration. The states with flux $\Phi$ in the central hexagon can be realized in the 6-site unit cell, while the states with flux $\Phi$ necessitate a 12 site primitive unit cell ({\it i.e.} 24 site hexagonal unit cell).
\end{enumerate}

We use the notation $t$ instead of the $\chi$ to distinguish the hopping amplitudes set by hand in the variational approach from the solutions $\chi$ of the mean-field approach. Since the lattice is bipartite, only the relative sign of the hoppings $t_h$ and $t_d$ is important, so we can choose $t_h>0$ and parametrize our results by the ratio $t_d/t_h$. Note that changing the sign of the ratio $t_d/t_h$ allows us to introduce an additional flux $\pi$ into the hexagons surrounding the central hexagon. Except for the chiral phases, both $t_h$ and $t_d$ can be chosen to be real numbers.

\begin{figure}
\begin{center}
\includegraphics[width=7.5truecm]{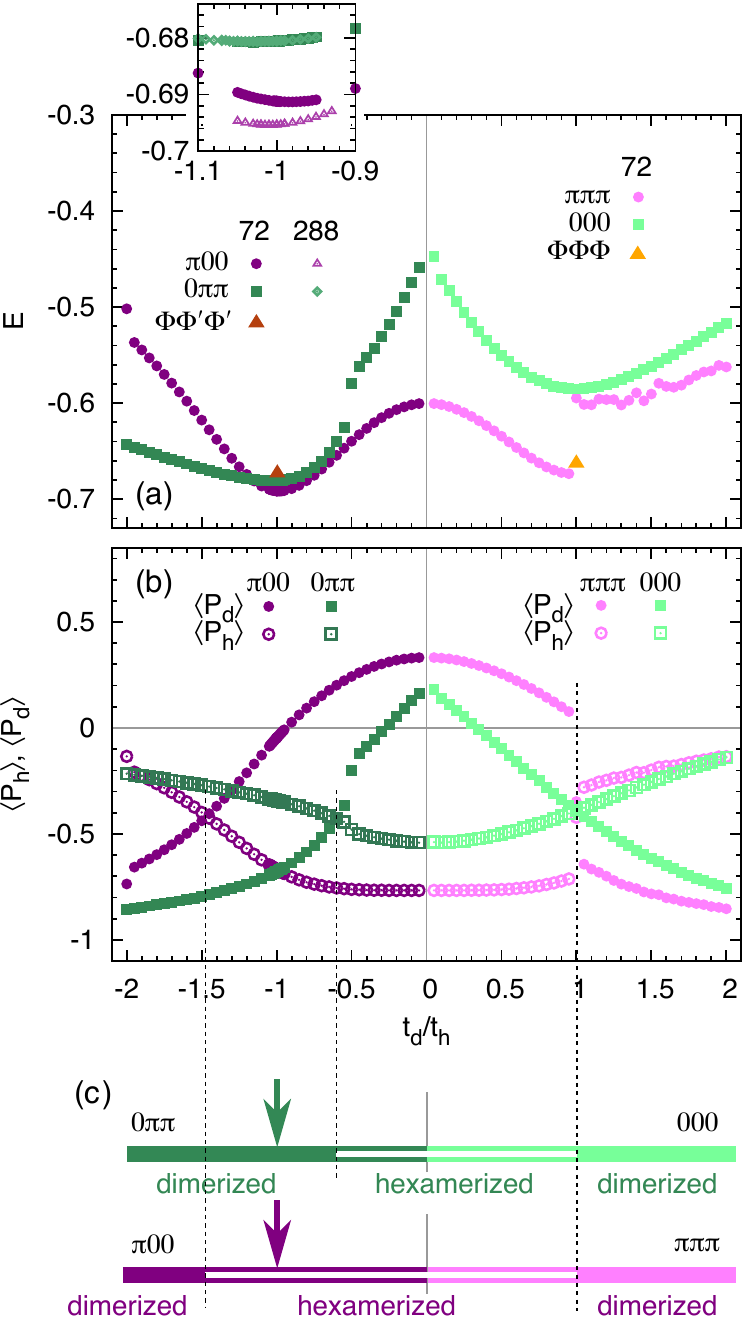} 
\caption{(Color online) 
(a) Nearest-neighbor bond energy as a function of $t_d/t_h$ in the 72 site cluster.  As we change the sign of $t_d/t_h$ from negative to positive, we shift between the $0\pi\pi$--flux and $000$--flux states (green squares), or between the $\pi00$--flux and $\pi\pi\pi$--flux states (purple circles). The energy of the chiral $\Phi\Phi\Phi$--flux state with $\Phi=2\pi/3$ is compared to the $t_d=t_h$ case, while the energy of the chiral $\Phi \Phi' \Phi'$--flux state is compared to the $t_d= -t_h$ case. The $\Phi' \Phi' \Phi'$--flux and $\Phi' \Phi \Phi$--flux states have a higher energy (not shown).  The inset of (a) shows the energies around $t_d/t_h=-1$ for the $72$ and $288$ site clusters. The ground state is degenerate for the $\pi\pi\pi$--flux state when $t_d/t_h>1$, and this is the origin of the scattered energy values.
(b) The energies of the $d$ and $h$ bonds versus $t_d$.  The hexamerization ($\langle P_h \rangle  <  \langle P_d \rangle $) is more extended for the $\pi00$--flux state than for for the $0\pi\pi$--flux state. (c) Schematic drawing of the extension of the hexamerized (plaquette) and dimerized phases that can be read off from the bond energies given in (b). The arrows denote the minima of the energies shown in (a). 
\label{fig:td_vs_E}
}
\end{center}
\end{figure}

\begin{figure*}
\begin{center}
\includegraphics[width=13truecm]{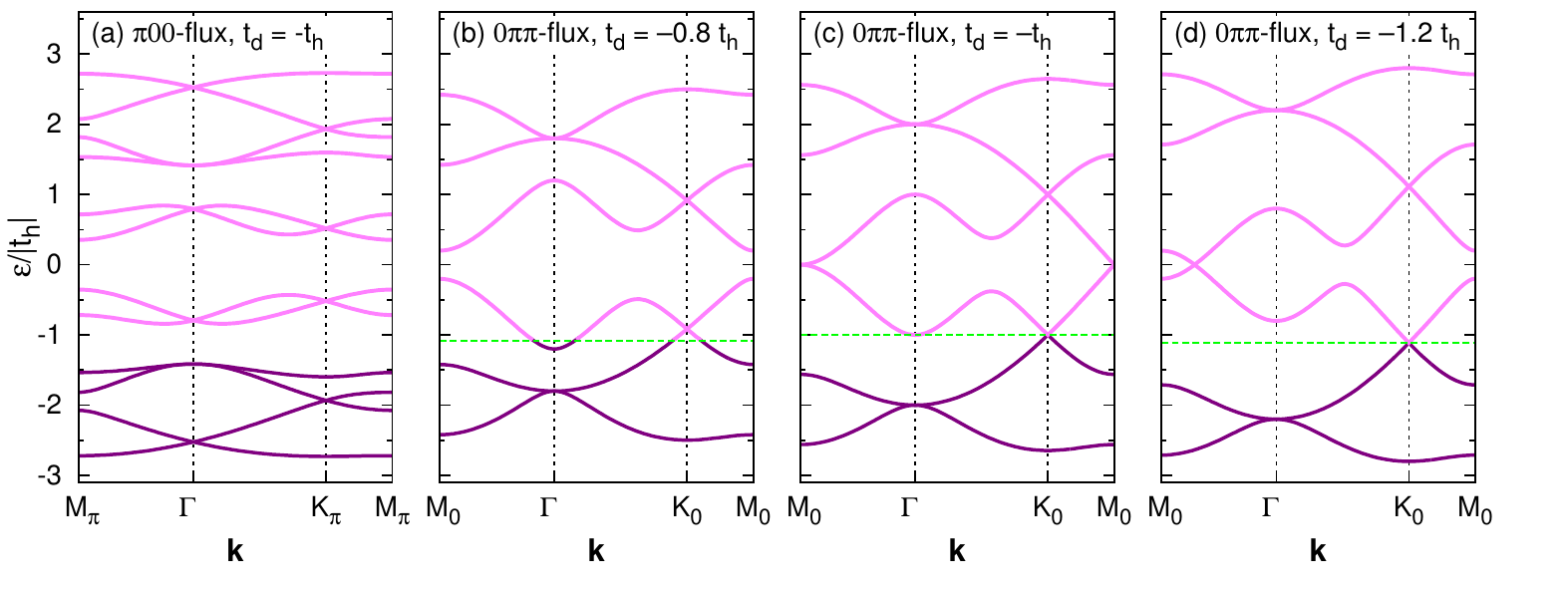} 
\caption{(Color online) 
The band structures of the free fermion Hamiltonian along the path $M_\pi\Gamma K_\pi M_\pi$ in the Brillouin--zone shown in Fig.~\ref{fig:6and24uc}(c)
 for   (a) the hexamerized $\pi00$-flux state ($t_d=-t_h$), (b)-(d) the dimerized $0\pi\pi$-flux state for different values of $t_d$. For $t_d>1$ the Fermi surface is a Dirac point at ${\mathsf K}_0$ --- the minimal energy of the corresponds to the case when the Fermi sea touches the $\varepsilon/|t_h| = -1$, $\Gamma$ point for $t_d = -t_h$ [plot (c)]. The dark and light purple lines denote occupied and empty bands, respectively, the green dashed line shows the Fermi energy.
\label{fig:bands}}
\end{center}
\end{figure*} 

\subsection{Variational Monte Carlo results}
In the following, we calculate the expectation value of the exchange Hamiltonian in the Gutzwiller projected wave function using Monte-Carlo importance sampling as a function of $t_d$ for the different realizations of the fluxes. We choose system sizes that have the full symmetry of the hexagonal lattice and that are compatible both with the 18-site unit cell of the SU(3) symmetry broken state with the long-range order shown in Fig.~\ref{fig:states}(a) and with the 24-site unit cell of the $\pi\pi\pi$ and $\pi00$ flux states. This leaves us with two families of clusters that have $72$, $2^2\times72=288$, $3^2\times72=648$,... and $216$, $2^2\times 216=864$,... sites. We have used the 72-site cluster to calculate the energy in a wide parameter range, and we have refined the calculation on the 288-site cluster around the minima found on the 72-site cluster. In our Monte-Carlo sampling an elementary update was the exchange of two randomly chosen fermions at arbitrary sites with different colors. Each run had $10^{10}$ ($2\times10^{10}$) elementary updates for the 72 (288) size cluster. To avoid any autocorrelation effect, we performed measurements after every 1000 (10000) updates, which corresponds to about 5 times the autocorrelation time. We averaged over all the bonds when calculating the energy, and the errors of the energy/site values are of the order of $10^{-4}$, thus much smaller than the symbol sizes on the plots. Furthermore, the boundary conditions (periodic or antiperiodic) of the hopping hamiltonian were chosen to avoid the degeneracy of the free--fermion ground state, if possible at all.

The Monte-Carlo results for the energy per site, given as $E= \langle P_h \rangle +  \langle P_d \rangle /2$, are 
shown in Fig.~\ref{fig:td_vs_E}(a), while  $\langle P_h \rangle$ and  $\langle P_d \rangle$, the expectation values of the exchange operator on the bonds of the hexagons and on the dimers, respectively, are shown in Fig.~\ref{fig:td_vs_E}(b). 
For $t_d=0$, the wave function is a product of decoupled hexagons. 
For a single hexagon with $\pi$--flux (antiperiodic boundary conditions), the variational calculation leads to $\langle P_h \rangle  = -23/30 = -0.7667$, very close to $\langle P_h \rangle  = -0.7676$ from exact diagonalization of a six site SU(3) Heisenberg chain. Since the ground state of a hexagon is a singlet, the exchange between the decoupled hexagons is $\langle P_d \rangle  = 1/3$. 
This is seen in  Fig.~\ref{fig:td_vs_E}, where the $\pi\pi\pi$ and $\pi00$ states meet at $t_d=0$. The $\pi$ phase in the hexagon with nonvanishing hopping amplitudes makes the free fermion levels twofold degenerate with $\varepsilon = \pm \sqrt{3} t_h$ and 0, so that we have a closed shell condition for two fermions, i.e. the two fermion ground state is non-degenerate with energy $-2\sqrt{3}t_h$ (we need to put 2 fermions of each color to reach 6 fermions per hexagon, corresponding to a filling of one fermion per site). By contrast, the energy levels of a  hexagon with $0$ flux are $\varepsilon = \pm 2t_h$ and $\varepsilon = \pm t_h$, the latter levels being twofold degenerate. In this case the two--fermion ground state is also twofold degenerate, with energy $-3t_h$. So, decoupled hexagons prefer to have $\pi$ flux. 
Hexamerization ($\langle P_h \rangle < \langle P_d \rangle$) is present for $0\leq t_d/t_h<1$ in both the  $\pi\pi\pi$ and $000$ phases, and dimerization takes over for $1<t_d/t_h$. In the $0\pi\pi$ state the hexamerization persists in the $-0.5 \alt t_d/t_h \leq 0 $ region, while for the $\pi00$ case the hexamerization is present in a larger window, for  $-1.5 \alt t_d/t_h \leq 0 $. For $t_h \rightarrow 0$ the value of $\langle P_d \rangle$ tends to $-1$.

Among all states that we have investigated, the $\pi00$-flux configuration provides the lowest energy per site, -0.6912 on the 72-site cluster. The minimum occurs around $t_d/t_h  \approx -1$, where the projected state shows strong hexamerization  (the $\langle P_h \rangle$ is significantly smaller than $ \langle P_d \rangle$, see Fig.~\ref{fig:td_vs_E}). The minimum of the $0\pi\pi$--flux wave function showing dimerization also occurs around $t_d/t_h \approx -1$, with an energy  -0.6807 per site ($N_s=72$) that is 
 slightly higher than that of the hexamerized $\pi00$-flux state. Furthermore, the energy of the dimerized state depends only weakly  on the cluster size, whereas the energy of the hexamerized state is significantly lowered when going to the 288-site cluster [cf. inset in Fig.~\ref{fig:td_vs_E}(b)]. This shows that in the thermodynamic limit the hexamerized $\pi00$-flux state has clearly a lower variational energy than the dimerized $0\pi\pi$--flux state, in agreement with the iPEPS results. The energy of the $\pi\pi\pi$, $000$, and chiral states are all above the $\pi00$ and $0\pi\pi$ states.\footnote{We have also compared the energy of different chiral states on the 72 site cluster:  
$E(\Phi\Phi'\Phi') = -0.671$,
$E(\Phi\Phi\Phi) = -0.662$,
$E(\Phi'\Phi\Phi) = -0.629$, and
$E(\Phi'\Phi'\Phi') = -0.604$, where $\Phi=2\pi/3$ and $\Phi'=\pi+\Phi=5\pi/3.$}

Next, let us investigate some additional properties of the ground state, the hexamerized $\pi00$--flux state, and of its main competitor, the dimerized $0\pi\pi$--flux state.
It is quite interesting that the minimum of the energy is around $t_d=-t_h$ in both cases. We have no explanation why this is so for the 
hexamerized $\pi00$--flux state since there is nothing particular happening at that point in the free fermion band structure (Fig.~\ref{fig:bands}(a)), 
the occupied bands being well separated from the unoccupied bands. By contrast, the Fermi level for the dimerized $0\pi\pi$--flux state is inside the bands (Fig.~\ref{fig:bands}(b)-(d)), and the $t_d=-t_h$ point is a special one where the Fermi energy sits both at a band edge and at a Dirac point: it separates a fermionic state with a finite Fermi surface from a state where the Fermi surface reduces to a high symmetry point $K_0$, at which there is a Dirac-cone.

%
%
The differences between the bond-energies on the dimers and hexamers for the minimal energy $\pi00$--flux state in the VMC is $\langle P_d \rangle - \langle P_h \rangle \approx  0.6-0.7$, i.e. larger than the iPEPS result of $\approx 0.3-0.45$.
For the $0\pi\pi$--flux dimerized case the VMC result is $\approx -0.35(5)$, while  iPEPS provides $\approx -0.26(2)$.

\begin{figure}
\begin{center}
\includegraphics[width=8truecm]{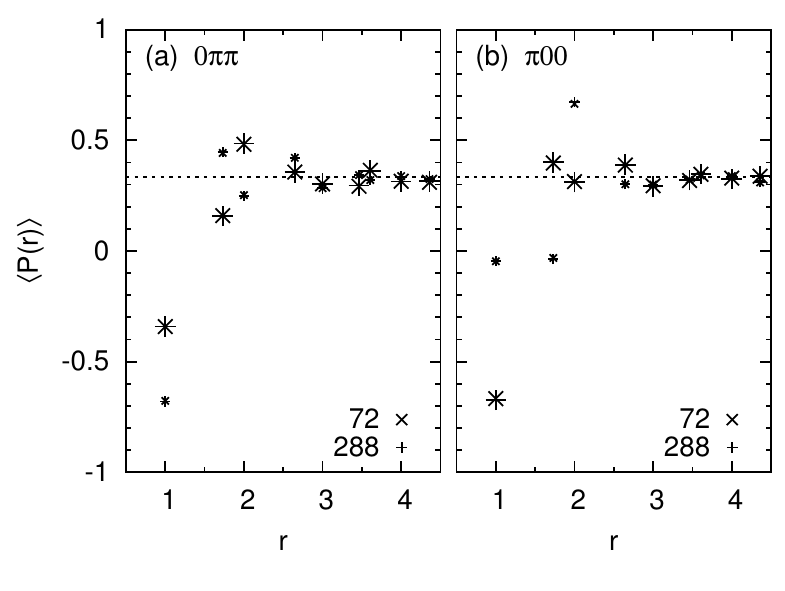}
\caption{
\label{fig:ninj}
 The expectation value of the $\langle P(r) \rangle = \langle P_{ij} \rangle$ operator, where $r$ is the distance between the sites $i$ and $j$, for the (a) dimerized $0\pi\pi$ and (b) hexamerized $\pi00$--flux phase, for the 72 and 288 site clusters with $t_d = -t_h$. The size of the symbols is proportional to the number of bonds having that value of $\langle P(r) \rangle$. For $r \rightarrow \infty$  the value of $\langle P(r)$ tends to $1/3$ (denoted by the thin black line), corresponding to the expectation value for the exchange between independent spins, which shows the absence of the long-range order.
}
\end{center}
\end{figure}

The color--color correlations $\propto \langle P_{ij} \rangle -1/3$ decay rapidly with the distance, as shown in Fig.~\ref{fig:ninj}.
Since there is a Dirac-cone in the free-fermion spectrum of the dimerized $0\pi\pi$--flux state for $t_d/t_h <-1$, the question whether the correlations decay algebraically, and not exponentially as expected for a gapped state, is legitimate.  Unfortunately, we cannot determine unambiguously the nature of the color-color correlations in real space, even when using the results from the 648-site cluster, the largest we considered. 

\begin{figure}
\begin{center}
\includegraphics[width=7truecm]{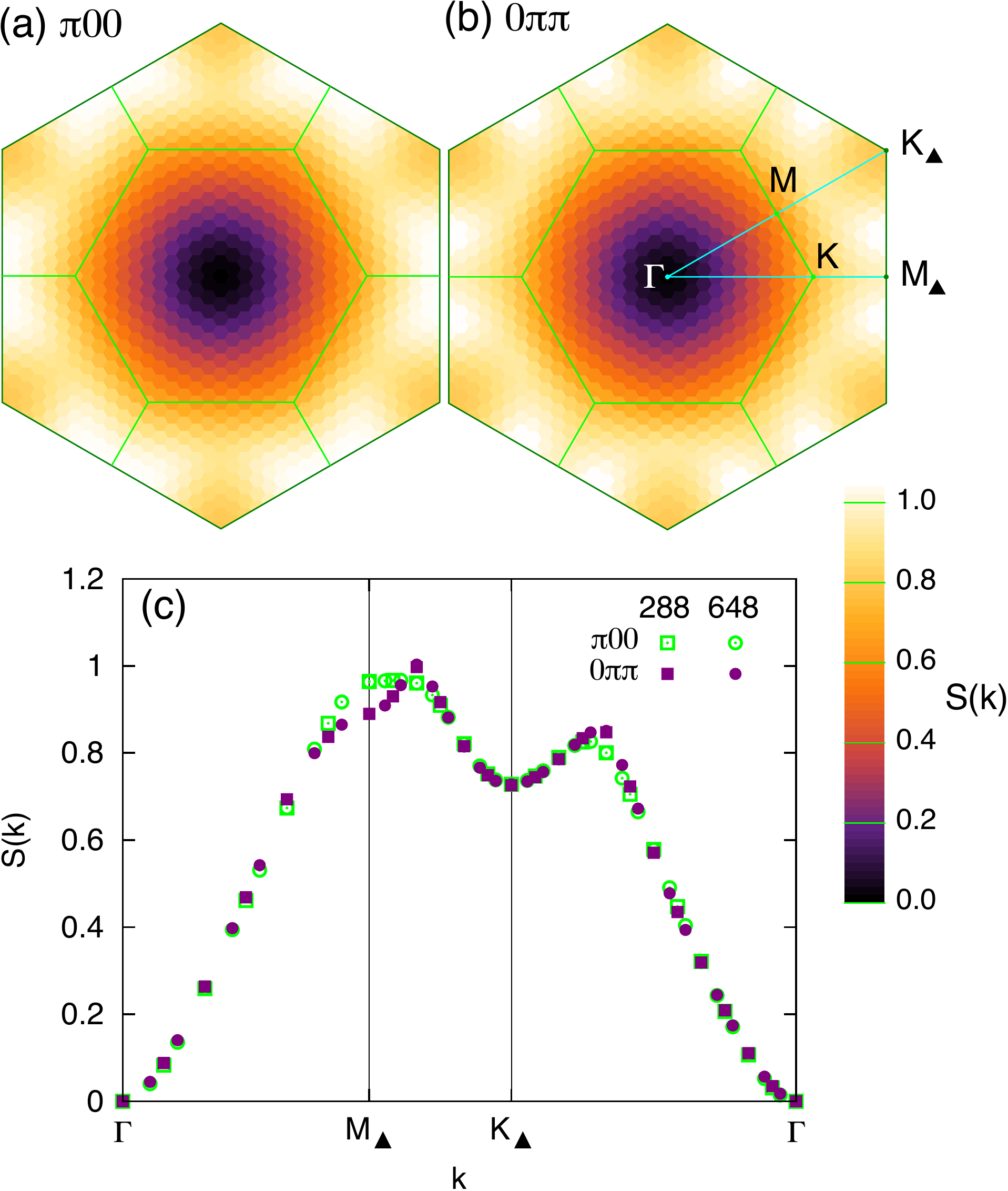}
\caption{\label{fig:sq}
(Color online) The spin structure factor $S({\bf k})$ in the $k$-space for the (a) hexamerized $\pi00$--flux and (b) dimerized $0\pi\pi$--flux state, calculated from the 648-site cluster, with $t_d=-t_h$. The light--green hexagons indicate the Brillouin zone of the honeycomb lattice, with the high-symmetry points $\Gamma$, $K$, and $M$. The $S({\bf k})$ for the $0\pi\pi$--flux state is peaked one--third way on the line between the points $M_\blacktriangle$ and $K_\blacktriangle$ in the second Brillouin--zone of the honeycomb lattice, as can also be seen in (c), where the $S({\bf k})$ is shown along the path $\Gamma M_\blacktriangle K_\blacktriangle \Gamma$ for the 288 and 648-site cluster. The $S({\bf k})$ for the $\pi00$--flux phase (open symbols), also with $t_d=-t_h$, shows no singular behavior. The spin structure factors for the two flux states are nearly identical away from the $M_\blacktriangle$ point.}
\end{center}
\end{figure}

Next, we look for the signature of the Dirac point in the structure factor 
\begin{equation}
  S({\bf k}) = \frac{1}{4} \sum_{j=1}^{N_{s}} 
   \left( \langle P_{0,j} \rangle -\frac{1}{3} \right) \cos {\bf k}\cdot ({\bf r}_{j}-{\bf r}_{0}) \;,
  \label{eq:sq}
\end{equation}
where the summation is over the $N_s$ sites of the cluster, and ${\bf r}_{i}$ is the position of site $i$. The prefactor is chosen so that the $\sum_{{\bf k}\in \text{BZ}_\blacktriangle} S({\bf k}) = N_s$ sum rule is satisfied, where the sum is over the $3 N_s/2$ ${\bf k}$ vectors of the Brillouin zone (with the high symmetry points $M_\blacktriangle$ and $K_\blacktriangle$) of the underlying triangular lattice which, in addition to the sites of the honeycomb lattice, also includes the centers of the hexagons. We find that the behavior of the structure factor $S({\bf k})$ is remarkably different for the $0\pi\pi$ and $\pi00$ cases [Fig.~\ref{fig:sq}] close to the $M_\blacktriangle$-point: in the former one $S({\bf k})$ is peaked in the second Brillouin zone, while the latter one is smooth. The position of the peak, when folded back to the Brillouin-zone of the 6--site unit cell, is at $K_0$, the wave vector that corresponds to the scattering between the Dirac points. 

\begin{figure}
\begin{center}
\includegraphics[width=7truecm]{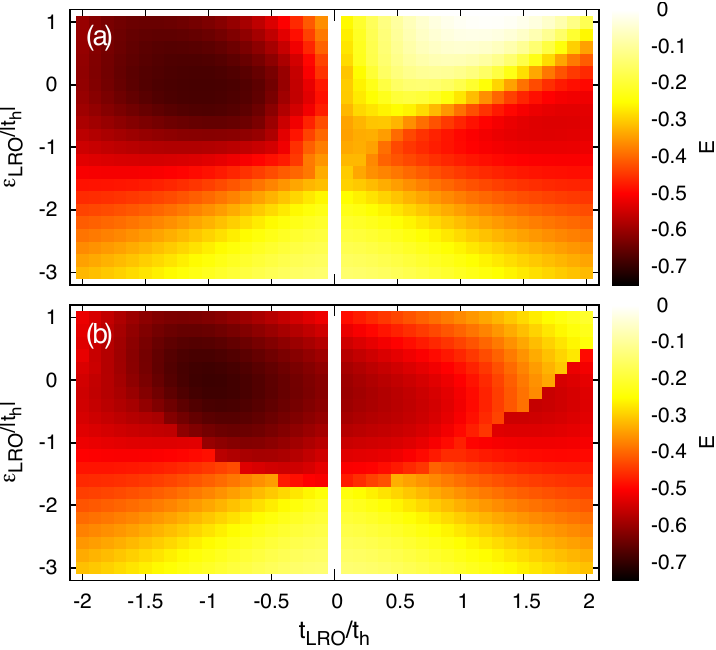}
\caption{(Color online) The stability of the dimerized and hexamerized states versus the formation of long range order, in the (a) $0\pi\pi$  and (b) $\pi00$ flux state, respectively. 
For both cases the energy is minimal for $\epsilon_{\text{LRO}}=0$ and $t_{\text{LRO}}=t_h$, where there is no long range order.  The calculations were made with  stepsizes $\Delta t_{\text{LRO}}= 0.1 |t_h|$ and $\Delta \epsilon_{\text{LRO}}=0.2 |t_h|$. For $t_{\text{LRO}}=0$ the fermionic band structure collapsed to a few highly degenerate bands, this caused difficulties in the Monte Carlo code we used for the calculation.}
\label{fig:2Denplot}
\end{center}
\end{figure}

\subsection{Check of the stability towards a color-ordered state}

Following the iPEPS results, which point to  a possible SU(N) symmetry breaking in the dimerized state, we further investigate the possibility of the formation of long-range order in the Gutzwiller-projected variational approach as well. 
To this end, we allow for color dependent hopping parameters and onsite energy terms, starting from the anticipated minimum energy variational state with $t_d=-t_h$. For each colored fermion we introduce a negative onsite energy $\epsilon_{\text{LRO}}$ according to the long--range ordered pattern of Fig.\ref{fig:statesiPEPS}(a) to enforce the SU(N) symmetry breaking in the Gutzwiller projected wave--function. Furthermore, we also allow for the modification of the hopping amplitude (denoted by $t_{\text{LRO}}$) for the colors that dominate the two sites of the bond, while leaving the hopping amplitude unchanged for the third color. The sign of the hopping parameters are set to display the appropriate flux state. Note that upon reversing the sign of $t_{\text{LRO}}$ the fluxes on the hexagons do not change since, for each color, we change the sign of two (or none) of the hopping parameters around a hexagon.
 Fig. \ref{fig:2Denplot} shows the energy of the Gutzwiller projected state as a function of $\epsilon_{\text{LRO}}$ and $t_{\text{LRO}}$ for the $0\pi\pi$ and $\pi00$ flux states.  It can be clearly seen that the energy is minimal for $\epsilon_{\text{LRO}}=0$ and $t_{\text{LRO}}=t_h$, i.e. for the case where there is no long-range order in the system. 
 We have repeated the calculation also for the $000$ and $\pi\pi\pi$ case as well, starting from $t_d=t_h$, and we found that the long-range color ordered phase is stabilized for the $000$--flux state with energy $E=-0.610$ per site (for $t_{\text{LRO}}/|t_H|=1.3$ and $\epsilon_{\text{LRO}}/|t_h|=-0.4$), much higher than the lowest energy $\pi00$ solution.


\section{Discussion and conclusion}
\label{sec:concl}
In this work we showed that the ground state of the SU(3) Heisenberg model on the honeycomb lattice has plaquette order and does not break SU(3) rotational symmetry in color space, in agreement with the previous tensor network study in Ref.~\onlinecite{Zhao12}. Extrapolations of the plaquette order parameter to the infinite $D$ limit reveals that the ground state indeed exhibits true long-range order. 

This result is in conflict with the prediction by LFWT from Ref.~\onlinecite{Lee12}. However, by performing a systematic study of the solution as a function of the bond dimension $D$ in iPEPS we can understand the LFWT prediction as a low-entanglement solution which is energetically favorable for small $D$, but which is unstable upon taking more quantum fluctuations into account by going to large $D$. 
This situation is reminiscent of the SU(4) Heisenberg model on the square lattice, where LFWT and iPEPS with $D=2$ predict a plaquette-color ordered state, however, for $D\ge5$ a dimer-N\'eel ordered state is stabilized.~\cite{corboz11-su4}

Thus, iPEPS is an ideal tool to compare competing states. Unlike standard variational methods, one can study how the individual energies of the competing states change upon increasing the amount of quantum fluctuations in a systematic way,  by varying the bond dimension. Such systematic analysis will be important also for future tensor network studies, e.g. for the $t$-$J$ model where a uniform d-wave superconducting state is competing with a \emph{striped} state at very low energies.~\cite{corboz2011,Hu12}


Next, comparing the energy of Gutzwiller projected wave functions, we have found that the competition between plaquette formation (hexamerization) and dimerization is also present in the variational treatment. We have identified 
that the two competing states originate from a free-fermion wave function with $\pi00$ and $0\pi\pi$ fluxes on the hexagons in the unit cell, with very different nature: the lower energy $\pi00$-flux state describes a gapped, plaquette (hexamerized) state, while the higher energy $0\pi\pi$ flux state is gapless and dimerized, with a free-fermion wave function having a Fermi-point at the Dirac cone. This difference is exemplified in the structure factor, in the former case it is a smooth function, in the latter it has a peak related to the position of the Dirac cones. A dimerized solution with a broken SU(3) symmetry has also been identified in the variational treatment which, however, has a much larger energy than the two competing states.

Finally we note that the plaquette ground state is very different from the one obtained for the same model on the square lattice, which exhibits three-sublattice N\'eel order.\cite{toth2010,Bauer12} In principle, length-6 loops could also be formed on the square lattice. However, the energy cost to introduce another weak bond (across the hexagons which results in a square lattice) is too high, which makes the three-sublattice N\'eel ordered state energetically favorable in this case.

\acknowledgments
The ED simulations have been performed on machines of the platform "Scientific computing" at the University of
Innsbruck - supported by the BMWF, and the iPEPS simulations on the Brutus cluster at ETH Zurich.
We thank the support of the Swiss National Science Foundation, MaNEP, the Hungarian OTKA Grant No. 106047
and the  Austrian Science Fund (FWF) through the SFB FoQuS (FWF Project No. F4006-N18)

\appendix

\bibliographystyle{apsrev4-1}
\bibliography{refs}

\end{document}